\pdfoutput=1

\documentclass[12pt,a4paper]{article}
\usepackage[utf8]{inputenc}
\usepackage{dsfont,amssymb} 
\usepackage{graphicx}
\usepackage[T1]{fontenc}
\usepackage{amsmath}
\usepackage{ctable}
\usepackage[colorlinks=true,linkcolor=blue]{hyperref}

\newcommand{\ie}{\emph{i.e.\/}}

\newcommand{\M}{\ensuremath{\mathds{M}}}
\newcommand{\R}{\ensuremath{\mathds{R}}}

\newcommand{\sidecaption}{\centering}
\title{Eigengestures for natural human computer interface}

\author{Piotr Gawron \and Przemys\l{}aw G\l{}omb \and Jaros\l{}aw Adam Miszczak \and Zbigniew Pucha\l{}a
\footnote{
Institute of Theoretical and Applied Informatics,
Polish Academy of Sciences, 
Ba\l{}\-ty\-cka 5, 44-100 Gliwice, Poland,
\texttt{\{gawron,przemg,miszczak,z.puchala\}@iitis.pl}
}
}
\date{May 6, 2011}
\begin{document}
\maketitle
\abstract{We present the application of Principal Component Analysis for data
acquired during the design of a natural gesture interface. We investigate the
concept of an eigengesture for motion capture hand gesture data and present the
visualisation of principal components obtained in the course of conducted
experiments. We also show the influence of dimensionality reduction on
reconstructed gesture data quality. 
}
\section{Introduction}
\label{sec:introduction}
Human-computer interface (HCI) which uses gestures promises to make certain
forms of user interfaces more effective and subjectively enjoyable. One of
important problems in creating such interface is the selection of gestures to
recognize in the system. It has been noted \cite{Wexelblat:1997} that choosing
gestures that are perceived by users as natural is one of decisive factors in
interface and interaction performance. At the same time, a large amount of
research is focused on fixed movements geared towards efficiency of recognition,
not interaction \cite{Wexelblat:1997}.

We view the analysis of natural gestures as a prerequisite of constructing an
effective gestural HCI. As a tool for this task, it is natural to use Principal
Component Analysis (PCA) \cite{jolliffePCA}. PCA has been successfully applied
for analysis and feature extraction \ie{} of faces (the famous `eigenface'
approach \cite{Turk:1991}). For human motion, PCA has been found to be a useful
tool for dimensionality reduction (see \ie{} \cite{witte09horseback}).
Eigengestures appear in a number of publications, \ie{} \cite{Nakajima:2006},
where they are used as input for motion predictor. In \cite{Yang:2007} they are
used for synthesis of additional training data for HMM. In \cite{Birk:1997}
eigengesture projection is used for real-time classification. We argue, however,
that the eigenvectors of human gestures--especially hand gestures--should be
investigated beyond the effect they have in improving data processing (\ie{}
classification score); the structure of the decomposition may lead to important
clues for data characteristics, as it has been the case for images
\cite{Hyvarinen:2009}. To the best of authors' knowledge, this is a still a
research field with limited number of contributions: in \cite{Yao:2010}
eigen-decomposition of 2D gesture images is only pictured without discussion,
whereas in \cite{Zhang:2010} a basic analysis is done only for whole body
gestures; main eigenvector are identified with deictic (pointing) gestures.

The main contribution of this work is application of PCA to analysis of the data
representing human hand gestures obtained using motion capture glove. We show
the influence of dimensionality reduction on reconstructed signal quality. 
We use the notion of eigengesture to the collected data in order to visualize
main features of natural human gestures.

This article is organized as follows. Section \ref{sec:method} presents the
experiment methodology; the sample set of gestures, acquisition methods,
participants and procedure. Section \ref{sec:dataprocessing} details the
application of PCA to motion capture gesture data. Section \ref{sec:application}
presents discussion the computed principal components. Section
\ref{sec:visualization} presents visualization of eigengestures. Last section
presents concluding remarks.
\section{Method}
\label{sec:method}
For our experiment, we used base of 22 different type of gestures, each type
represented by 20 instances --- 4 people performing the gestures, each of them
made the gesture 5 times (three with normal speed, then one fast following with
one slow execution). The gestures are detailed in table \ref{table:gestureList}.
For discussion on gesture choice the reader is referred to~\cite{Glomb:2011}.
\begin{center}
\newcounter{gestureCounter}\stepcounter{gestureCounter}
\newcommand{\gc}{\arabic{gestureCounter}\stepcounter{gestureCounter}}
\ctable[
  cap = Gesture list,
  caption = The gesture prepared with the proposed methodology,
  label = table:gestureList,
]{rllll}{
  \tnote{We use the terms `symbolic', `deictic', and `iconic' based on McNeill \& Levy \cite{McNeill:1992} classification, supplemented with a category of `manipulative' gestures (following \cite{Quek:2002})}
  \tnote[b]{Significant motion components: T-hand translation, R-hand rotation, F-individual finger movement}
  \tnote[c]{This gesture is usually accompanied with a specific object (deictic) reference}
}{\FL
& Name & Class\tmark & Motion\tmark[b] & Comments\ML
\gc & \emph{A-OK} & symbolic & F & common `okay' gesture\NN
\gc & \emph{Walking} & iconic & TF & fingers depict a walking person\NN
\gc & \emph{Cutting} & iconic & F & fingers portrait cutting a sheet of paper\NN
\gc & \emph{Shove away} & iconic & T & hand shoves away imaginary object\NN
\gc & \emph{Point at self} & deictic & RF & finger points at the user\NN
\gc & \emph{Thumbs up} & symbolic & RF & classic `thumbs up' gesture\NN
\gc & \emph{Crazy} & symbolic & TRF & symbolizes `a crazy person'\NN
\gc & \emph{Knocking} & iconic & RF & finger in knocking motion\NN
\gc & \emph{Cutthroat} & symbolic & TR & common taunting gesture\NN
\gc & \emph{Money} & symbolic & F & popular `money' sign\NN
\gc & \emph{Thumbs down} & symbolic & RF & classic `thumbs down' gesture\NN
\gc & \emph{Doubting} & symbolic & F & popular (Polish?) flippant `I doubt'\NN
\gc & \emph{Continue} & iconic\tmark[c] & R & circular hand motion `continue', `go on'\NN
\gc & \emph{Speaking} & iconic & F & hand portraits a speaking mouth\NN
\gc & \emph{Hello} & symbolic\tmark[c] & R & greeting gesture, waving hand motion\NN
\gc & \emph{Grasping} & manipulative & TF & grasping an object\NN
\gc & \emph{Scaling} & manipulative & F & finger movement depicts size change\NN
\gc & \emph{Rotating} & manipulative & R & hand rotation depicts object rotation\NN
\gc & \emph{Come here} & symbolic\tmark[c] & F & fingers waving; `come here'\NN
\gc & \emph{Telephone} & symbolic & TRF & popular `phone' depiction\NN
\gc & \emph{Go away} & symbolic\tmark[c] & F & fingers waving; `go away'\NN
\gc & \emph{Relocate} & deictic & TF & `put that there'\LL
}
\end{center}
The gestures were recorded with DG5VHand motion capture glove \cite{DG5VHand},
containing 5 finger bend sensors (resistance type), and three-axis accelerometer
producing three acceleration and two orientation readings. Sampling frequency
was approximately 33 Hz.

The participants for the experiments were chosen from the employees of Institute
of Theoretical and Applied Informatics of the Polish Academy of Sciences. Four
males were instructed which gestures were going to be performed in the
experiments and were given instructions how the gestures should be performed.
A~training session was conducted before the experiment.

During the experiment, each participant was sitting at the table with the motion
capture glove on his right hand. Before the start of the experiment, the hand of
the participant was placed on the table in a fixed initial position. At the
command given by the operator sitting in front of the participant, the
participant performed the gestures. Each gesture was performed three times at
the natural pace. Additionally, each gesture was made once at a rapid pace and
once at a slow pace. Gestures number 2, 3, 7, 8, 10, 12, 13, 14, 15, 17, 18, 19
are periodical and in their case the single performance consisted of three
periods. The operator decided about the end of data acquisition.

Input data consist of sequences of vectors $g_{t_n}\in \R^{10}, n\in\{1,N_i\}$
which are state vectors of the measurement device registered in subsequent
moments $t_n$ of time. The time difference $t_{n+1}-t_n$ is almost constant and
approximately 30 ms. Each registered gesture forms a matrix $G_i\in
M_{N_i,10}(\R).$ Acquisition time for every gesture is different, therefore the
number of samples $N_i$ depends on the sample. We acquired $K=22$ different
gestures, which are repeated $L=20$ times.
\section{Data processing}
\label{sec:dataprocessing}
Our chosen statistical tool was Principal Component Analysis (PCA). It has been
successfully applied in the domain of signal processing to various datasets
such as: human faces \cite{Turk:1991}, mesh animation \cite{Alexa:2000}.

\subsection{Principal Component Analysis}
For the sake of consistency we start by recalling basic facts concerning
Singular Value Decomposition (SVD)~\cite{vanloan} and Principal Component
Analysis (PCA)~\cite{jolliffePCA}.

Let $A\in \M_{m,n}$ has rank $k\leq m$. Then there exist orthogonal matrices 
$U\in \M_m$ and $V\in \M_n$ such that
$
A=U\Sigma V^T.
$
The matrix $\Sigma = \{\sigma_{ij}\} \in \M_{m,n}$ is such that
$ \sigma_{ij}=0,\ \mathrm{for}\ i\not=j,$
and
$
\sigma_{11}\geq\sigma_{22}\geq \ldots \geq\sigma_{kk}>\sigma_{k+1,k+1}
= \ldots = \sigma_{qq} = 0,
$
with $q=\min(m,n)$. 

The numbers $\sigma_{ii}\equiv \sigma_{i}$ are called \emph{singular values},
\ie{} non-negative square roots of the eigenvalues of $AA^T$. The columns of $U$
are eigenvectors of $AA^T$ and the columns of $V$ are eigenvectors of $A^T A$.

Principal Component Analysis allows us to convert a set of observations of
correlated variables into the so-called \emph{principal components}, \ie{} a set
of values of uncorrelated variables.

Formally, the $i$-th principal component is the $i$-th column vector of the
matrix $V_{:,i}\times \sigma_{ii}$ obtained as a SVD of the data matrix. In
order to perform PCA on the data acquired in different units, the data need to
be unified to a common units. In our case, the initial vector of data is
transformed by the \emph{studentisation}, \ie{} by subtracting the empirical
mean and dividing by the empirical standard deviation.

\subsection{Organisation of data}
As the input of the algorithm we have $K\times L$ matrices $G_i$. Each matrix
represents a~single realisation of a~gesture. In order to perform PCA, the data
are transformed in the following way:
\begin{enumerate}
    \item Re-sampling: for every signal indexed by $s\in \{1,\ldots, S=10\}$:
    $G_{t_n,s} \rightarrow G'_{t'_n,s},$ where $t_n$ indexes time samples of the
    gesture as acquired from the capture device, $t'_n\in\{1,\ldots, N=20\}$
    using linear interpolation.
    \item Arranging into the tensor: $$T_{k,l,t'_n,s}=(G'_{t'_n,s})_{k,l},$$ where
    $k\in \{1,\ldots, K=22\}$ denotes number of the gesture type and $l\in
    \{1,\ldots, L=10\}$ denotes individual realisation of a gesture.
    \item Double integration of signal from accelerometers to transform
    acceleration into position variable.
    \item Centring and normalisation: for every signal $s$: $T'_{k,l,:,s}=
    (T_{k,l,:,s}-\overline{T_{:,:,:,s}})/\sigma(T_{:,:,:,s})$.
\end{enumerate}

The data are arranged into a matrix $X_{(k,l),(t_n,s)}=T'_{k,l,t_n,s}$ whose
columns consists of vectorised distinct realisations of gestures.
Such a matrix is then feed into SVD algorithm.

\begin{figure}[ht]
    \sidecaption
    \includegraphics[width=\textwidth]{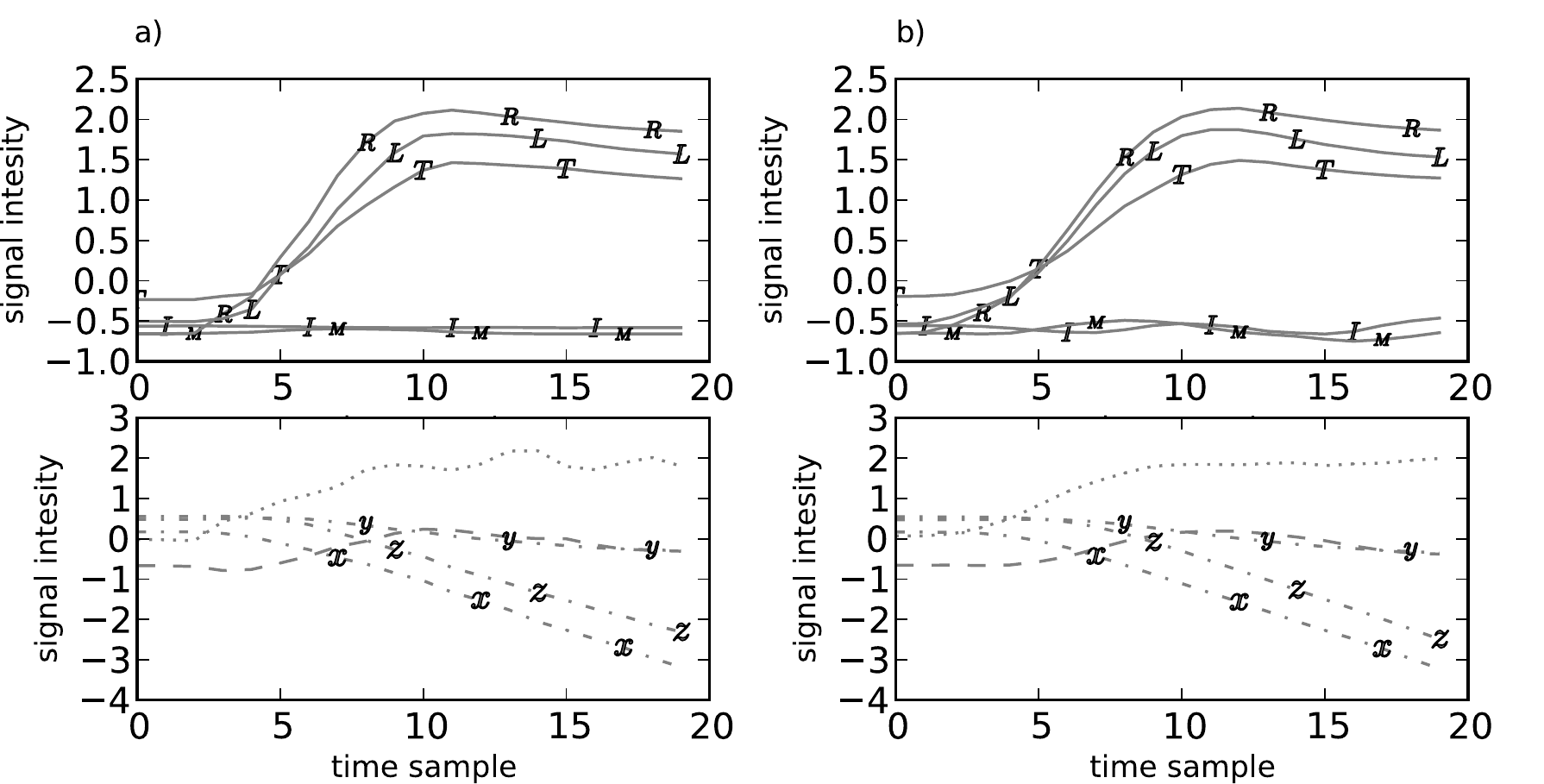}
    \caption{A sample of the gestures dataset. The data are normalised and
    centred. Single realisation of \emph{Cutting} gesture. Upper plot bending
    of fingers: $T$ --- thumb, $I$ --- index, $M$ --- middle, $R$ --- ring, $L$
    --- little; lower plot: dashed line --- palm roll, dotted line --- palm
    pitch, $X$ $Y$ $Z$ --- palm position in space. a) original data, b)
    approximation reconstructed using only 20 first principal components.}
    \label{fig:normalizedgesture}
\end{figure}

A sample of our data is visualised in Fig.~\ref{fig:normalizedgesture}
a) which presents the re-sampled, centred and rescaled second realisation of the
\emph{Cutting} gesture described in our data tensor by sub-matrix
$T'_{3,2,:,:}.$

\section{Application of PCA to data ex\-plo\-ra\-tion}
\label{sec:application}
\begin{figure}[ht]
\sidecaption
\includegraphics[width=0.45\textwidth]{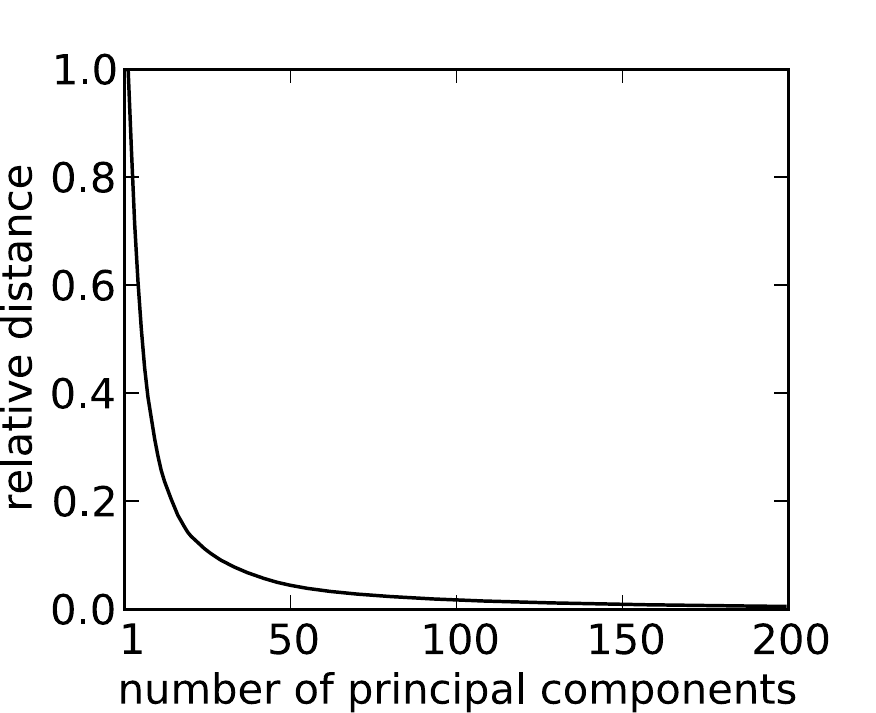}
\caption{Relative Euclidean distance between the dataset and its approximation
obtained using first $n$ principal components. }
\label{fig:approx}
\end{figure}

One of the typical applications of PCA to the analysis of the data obtained from
the experiment is to reduce their dimensionality. Fig.~\ref{fig:approx} shows
mean quality of the approximation of the original dataset in function of the
number of principal components used to reconstruct the dataset. The distance in
the Figure is scaled so that the approximation using only the first principal
component gives $1$. It can be easily seen that the dataset can be efficiently
approximated using low rank approximation. 

A comparison of original data sample {\it vs} its low rank approximation is
shown In Fig.~\ref{fig:normalizedgesture}, sub-plot a) shows original data for
\emph{Cutting} gesture and sub-plot b) shows the same data reconstructed using
only first 20 principal components.
\section{Visualization of eigengestures}
\label{sec:visualization}
The coordinates in which the eigengestures are obtained are artificial. To
create a visualisation one needs to change the coordinates to suited for the
\emph{hand presentation model}.

The change of coordinates is obtained by affine transformation acting
independently on each dimension (sensor data). The parameters of these
transformations (scales and translations) are obtained in the following way:
\begin{itemize}
\item The scale factor for each sensor (dimension) is a quotient of $0.05$ --
$0.95$ quantile dispersion of this sensor data and the dispersion of the sensor
in the eigengesture.
\item The translation is calculated in such way that each visualised
eigengesture has unified starting position.
\end{itemize}

\begin{figure}
\centering
\includegraphics[width=0.75\textwidth]{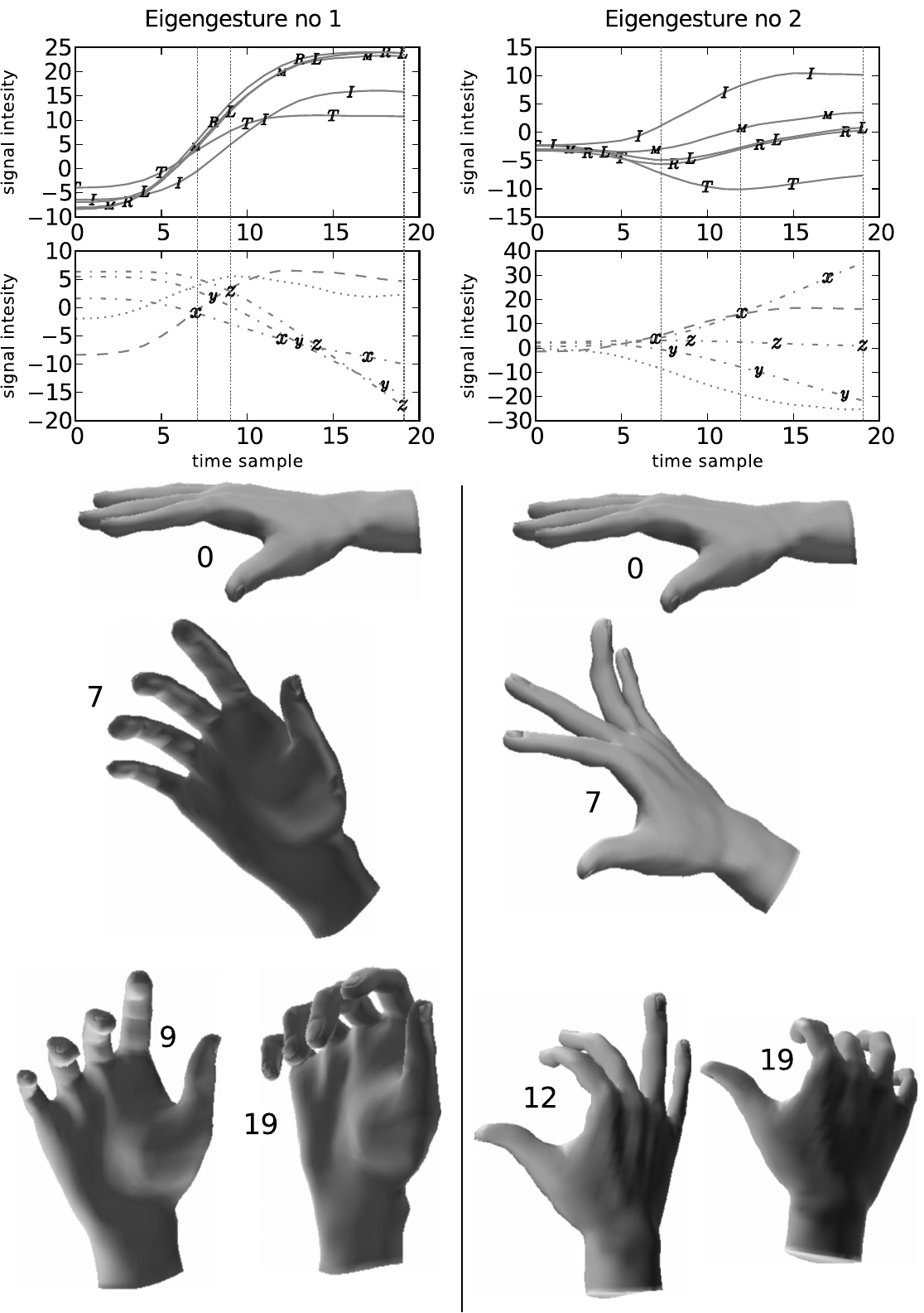}
\caption{Visualization of two first eigengestures (principal components). On top:
normalized and centred plots of signals in time. Upper plot bending of fingers:
$T$ --- thumb, $I$ --- index, $M$ --- middle, $R$ --- ring, $L$ --- little;
lower plot: dashed line --- palm roll, dotted line --- palm pitch, $X$ $Y$ $Z$
--- palm position in space. At the bottom: shapes of hands in selected time
moments. View is from the perspective of a person performing the gesture. 
For the sake of the clarity of the picture space position of the palm is 
omitted.
}
\label{fig:eigengestures}
\end{figure}

In Fig.~\ref{fig:eigengestures} the first two eigengestures (principal
components) are shown. The first eigengesture looks very natural and resembles
gestures commonly used by humans in process of communication. We found that
higher eigengestures do not look very natural especially because of the negative
values of the finger bends. Due to orthogonality of left singular vectors
obtained from SVD it can not be expected that eigengestures will be similar to
natural gestures performed by humans.

We observed that time plots of eigengestures around number 100 and higher are
very noisy.

\section{Conclusions and future work}
\label{sec:conclusions}
In this work our goal was to explore the space of human gestures using Principal
Component Analysis. Visualisation of eigengestures is a~tool that allows us to
understand better the dataset we acquired during the experiment. We have shown 
that natural human gestures acquired with use of motion caption device can be 
efficiently approximated using 50 to 100 coefficients. Additionally we have
identified the principal component of the gestures dataset.

Future work will consists of application of the obtained results to analysis of
quality of gesture recognition. We will compare PCA with Higher Order Singular
Value Decomposition \cite{Kolda:2008} as data dimensionality reduction techniques.

\section*{Acknowledgements}
This work has been partially supported by the Polish Ministry of Science and
Higher Education projects NN516405137 and NN519442339.
We would like to thank Pawe\l{} Kowalski and Sebastian Opozda for providing and
adapting the model of the hand. We would like to thank Dr Ryszard Winiarczyk for
encouraging us to publish this work.

The final publication is available at www.springerlink.com.

\end{document}